\documentstyle[prd,aps,psfig]{revtex}
\draft
\begin{document}
\title{
Instability of Quark Matter Core in a Compact Newborn Neutron Star With 
Moderately Strong Magnetic Field
}

\author{
Sutapa Ghosh$^{a)}${\thanks{E-mail:sutapa@klyuniv.ernet.in}} and  Somenath 
Chakrabarty$^{a),b)}${\thanks{E-mail:somenath@klyuniv.ernet.in}}
}

\address{
$^{a)}$Department of Physics, University of Kalyani, Kalyani 741 235,
India and
$^{b)}$Inter-University Centre for Astronomy and Astrophysics, Post Bag 4,
Ganeshkhind, Pune 411 007, India
}

\date{\today}
\maketitle

\begin{abstract}
It is explicitly shown  that if phase transition occurs at the core of a
newborn neutron star with moderately strong magnetic field strength,
which populates only the electron's Landau levels,  then
in the $\beta$-equilibrium condition, the quark core is energetically
much more unstable than the neutron matter of identical physical condition.
\end{abstract}
One of the oldest subject-"the effect of strong magnetic field on dense
matter" has gotten a new life after the discovery of a few strongly
magnetized neutron stars- which are called magnetars \cite{R1}. These
exotic stellar
objects are also assumed to be the possible sources of soft gamma
repeaters (SGR) and anomalous X-ray pulsars (AXP) \cite{R2,R3,R4}. 
From observations, the surface magnetic
field of such objects are found to be $\sim 10^{15}$Gauss \cite{R5}. 
The field at
the core region is  expected to be a few orders of magnitude larger than 
the surface
value. But there is an upper limit for magnetic field strength, beyond 
which the core
region of the star becomes unstable \cite{R6}. This value is $\approx
10^{18}$Gauss. The magnetars are also thought to be the strongly magnetized
young neutron stars. The studies on the effect of strong magnetic field
on various physical processes,
relevant for these exotic objects have been reported during the past few years.
These studies are mainly related to the equation of states of dense
matter \cite{R7,R8,R9},
elementary processes- specially weak and electromagnetic decays and
reactions \cite{R10}, quark-hadron phase transition \cite{R11,R12,R13}, and 
transport coefficients of 
dense matter \cite{R14}.  A few years ago we have shown explicitly that a first
order quark-hadron phase transition is absolutely forbidden in presence
of strong magnetic field ($\geq$ a few times $10^{15}$Gauss
\cite{R11,R12}). However,
in a recent paper, Mathews et. al. have shown \cite{R15} that such  strong
conclusion is not correct if one considers anomalous magnetic moment of quarks
in the quark matter sector. In that case a first order quark-hadron phase 
transition is possible even if the magnetic field is extremely
strong.  In the same publication we have also shown with certain
approximation, that even if a phase transition occurs at the core region
of a compact neutron star in presence of strong magnetic fields, in the
$\beta$-equilibrium condition the matter becomes energetically unstable
compared to neutron matter of identical physical condition \cite{R12}.
Hence we concluded that quark matter core is impossible in a strongly
magnetized young neutron star. In this
brief report we shall show explicitly without any approximation, that
the conclusion is still valid if the magnetic field strength is  
moderately strong. If it is correct, then
we can  very strongly conclude that the quark matter is absolutely
impossible at the core of a neutron star with magnetic field strength
slightly greater than $4.4\times 10^{13}$Gauss, which is the quantum
mechanical limit for electrons to populate Landau levels. We believe
that such a  conclusion is extremely important
both from the theoretical as well as observational points of view. 

In this report we have considered a young compact neutron star with
moderately high magnetic field (we consider a field strength of
$10^{14}$Gauss at the core region for our calculation). The density of
the core region is assumed to be such that a quark-hadron phase
transition (which is assumed to be first order even without the inclusion 
of anomalous magnetic dipole moment of quarks) 
can occur. Now in this case the assumed magnetic field strength at the core is
about a factor of two larger than the above critical value to populate Landau
levels for electrons. Further, the magnitude is not too high to affect
quantum mechanically other charged components present in the system
(e.g., $u$, $d$ and $s$ quarks) or populate only the zeroth Landau level
for electron. We have noticed that under such circumstance, an exact
estimation of the rates of weak processes are possible. Therefore in our
opinion, the uncertainty present in our previous publication is removed
in the range of magnetic field strength $B_c^{(e)} < B < B_C^{(u,d)}$
\cite{R13}. 

Now it is known that quark-hadron phase transition is a strong
interaction phenomenon, and therefore takes place in the strong interaction
time scale. On the other hand, immediately after phase transition, the
nascent quark matter is not necessarily in $\beta$-equilibrium 
configuration. This
is achieved through weak processes in the weak interaction time scale,
which is several orders of magnitude larger than the strong interaction
time scale. The goal of the present report is to show that if
quark-hadron phase transition occurs at the core of a moderately strong
magnetic field, so that only electrons at the back ground are affected
quantum mechanically,
then in the $\beta$-equilibrium condition, quark
matter phase becomes energetically much more unstable than the
corresponding neutron matter state.

To investigate the instability of quark matter core, we solve
numerically the set of
kinetic equations  for the nascent quark phase, which ultimately 
lead to chemical equilibrium configuration. We have considered the most
simplified physical picture in the quark matter sector-quarks are
non-interacting, at the very beginning, quark-hadron phase transition
occurred from non-strange hadronic matter and neutrinos are
non-degenerate- they leave the system immediately after their creation.
The relevant weak process are:
$d\rightarrow u+e^- +\bar{\nu_e}~(1),~ u+e^- \rightarrow d+\nu_e ~(2),~ 
s\rightarrow u+e^- +\bar{\nu_e} ~(3),~ u+e^- \rightarrow s+\nu_e ~ (4),  
~ u+d \leftrightarrow u+s ~(5).$
The approach to chemical equilibrium is governed by the following sets
of kinetic equations:
\begin{eqnarray}
\frac{dY_u}{dt}&=& \frac{1}{n_B}[\Gamma_1- \Gamma_2+\Gamma_3-\Gamma_4]\\
\frac{dY_d}{dt}&=& \frac{1}{n_B}[-\Gamma_1+ \Gamma_2-\Gamma_5^{(d)}+
\Gamma_5^{((r)}]
\end{eqnarray}
where $n_B$ is the baryon number density, $Y_i=n_i/n_B$ is the fractional 
abundance of the species $i$ and $\Gamma_j$'s are the rates of the 
processes $j=1,2,3,4,5$. The indices $d$ and $r$ are respectively for the 
direct and reverse processes for  $j=5$.
The baryon number conservation and charge neutrality conditions  
give $Y_s=3-Y_u-Y_d$ and $Y_e=Y_u-1$ respectively.
To solve the kinetic equations numerically for the investigation of  chemical 
evolution, we use these constraints as subsidiary conditions to obtain
$Y_e$ and $Y_s$, and
further we use the numerical values for the rates $\Gamma_1$ to
$\Gamma_5^{(d)(r)}$ appear on the right hand sides. Now for
a neutron star of mass $\approx 1.4M_\odot$, the baryon number density at 
the centre is $3-4$ times normal nuclear density, temperature $\sim 10^9$K and 
proton fraction is about $4\%$. Then the initial conditions are $Y_u(t=0)=1.04$,
$Y_d(t=0)=1.96$. As a consequence of baryon number conservation and charge 
neutrality, we have $Y_s(t=0)=0$ and $Y_e(t=0)=0.04$.

Since the magnetic field strength is assumed to be $\approx
10^{14}$Gauss, the rates for the first
four processes will be affected through electron spinor solution and
energy eigen value.
Further,  the rates for the processes (3) and (4) can very easily be
obtained from the rates of processes one and two respectively just by replacing
$d$-quark parameters with the corresponding $s$-quark ones and $\cos
\theta_c$ by $\sin \theta_c$, where $\theta_c$ is the well known Cabibbo
angle.

Now from the definition, the transition matrix element for the weak decay 
processes is given by
\begin{equation}
T_{fi}=\frac{4iG}{\sqrt 2} \cos \theta_c \int d^4x \left [\bar
\psi_u(x)\gamma_\mu \frac{1-\gamma_5}{2} \psi_d(x) \right ]
\left [\psi_e(x)\gamma^\mu \frac{1-\gamma^5}{2} \psi_\nu(x) \right ]
\end{equation}
Then the decay rate is given by
$d\Gamma = \lim_{\tau \rightarrow \infty} \mid T_{fi}\mid^2
d\rho_f/\tau$
where $\tau$ is the characteristic collision time and $\rho_f$ is the
final density of states, given by
$d\rho_f=\prod_i d^3p_i/(2\varepsilon_i(2\pi)^3)$,
where the product is over all final states $i$ and $\varepsilon_i$
is  single particle energy of the $i$th component. We have designated
$d,$ $\nu_e$ or $\bar \nu_e$, $u$ and $e$ by $i=1,2,3$ and $4$ respectively. 
In this moderately strong magnetic field strength, we have used conventional
spinor solutions for the quarks and charge neutral neutrinos or
anti-neutrinos, whereas for electron we have used \cite{R11,R13,R16}
\begin{equation}
\Psi^{(\uparrow)}(x)=\frac{1}{\sqrt {L_yL_z}} 
\frac{\exp(-i\varepsilon_{\nu}^{(i)}t+ip_y y+ip_z z)}
{[2\varepsilon_{\nu}^{(i)}(\varepsilon_{\nu}^{(i)}+m_i)]^{1/2}}
\left( \begin{array}{c}(\varepsilon_{\nu}^{(i)}+m_i) I_{\nu;p_y}(x)\\
0\\p_z I_{\nu;p_y}(x)\\-i(2\nu q_i B_m)^{1/2}I_{\nu-1;p_y}(x)
\end{array} \right)
\end{equation}
and
\begin{equation}
\Psi^{(\downarrow)}(x)=\frac{1}{\sqrt {L_yL_z}}
\frac{\exp(-i\varepsilon_{\nu}^{(i)}t+ip_y y+ip_z z)}
{[2\varepsilon_{\nu}^{(i)}(\varepsilon_{\nu}^{(i)}+m_i)]^{1/2}}
\left( \begin{array}{c} 0\\(\varepsilon_{\nu}^{(i)}+m_i)I_{\nu-1;p_y}(x)\\
i(2\nu q_i B_m)^{1/2}I_{\nu;p_y}(x)\\-p_z I_{\nu-1;p_y}(x)   
\end{array} \right)
\end{equation}
where the symbols $\uparrow$ and $\downarrow$ are used for up and down spin 
states respectively and
\begin{equation}
I_{\nu;p_y}(x)=\left( \frac{q_i B_m}{\pi}\right)^{1/4}
\frac{1}{\sqrt{\nu!}~2^{\nu/2}}
\exp\left[ -\frac{1}{2}q_i B_m \left( x- \frac{p_y}{q_i B_m}\right)^2 \right]
H_{\nu} \left[ \sqrt{q_i B_m}\left( x- \frac{p_y}{q_i B_m} \right) \right],
\end{equation}
$H_{\nu}$ is the well known Hermite polynomial of order $\nu$.
Then we have 
\begin{equation}
T_{fi}=-\frac{iG ~ 2\pi \delta(\varepsilon_1-\varepsilon_2\varepsilon_3
-\varepsilon_4)}{\sqrt{2}V^{3/2}}\Pi^\prime
\end{equation}
where
\begin{equation}
\Pi^\prime=[\bar u(p_3)\gamma_\mu (1-\gamma_5)u(p_1)][\bar
f_e(p_4)\gamma^\mu (1-\gamma^5)v(p_2)]\cos \theta_c,
\end{equation}
\begin{equation}
\bar f_e(p_4)=\int d^3x \exp [-i(\vec p_1 -\vec p_2 -\vec p_3).\vec r]
\bar \psi_e(x)
\end{equation}
Hence we have
\begin{eqnarray}
T_{fi}&=& -\frac{iG}{\sqrt{2}} \cos \theta_c (2\pi)^3
\frac{\delta(\varepsilon_1 -\varepsilon_2- \varepsilon_3-
\varepsilon_4)}{V^{3/2}} \nonumber \\
&& \delta(p_{1y}-p_{2y}-p_{3y}-p_{4y}) \delta(p_{1z}-p_{2z}-p_{3z}-p_{4z})
\Pi
\end{eqnarray}
where
\begin{equation}
\Pi= [\bar u(p_3)\gamma_\mu (1-\gamma_5)u(p_1)][\bar
u(p_4)\gamma^\mu (1-\gamma^5)v(p_2)]
\end{equation}
and
\begin{equation}
\bar u(p_4)^{(\uparrow)(\downarrow)}=\int \frac{dx}{\sqrt{L_yL_z}} \exp
[i(p_{1x}-p_{2x}-p_{3x}).x] (\uparrow)(\downarrow)
\end{equation}
where the symbols $(\uparrow)$ and $(\downarrow)$ indicate positive
energy up and down spin states for electron. Now to obtain
$\bar u(p_4)$ we have  evaluated 
\begin{equation}
\int_\infty^\infty dx \exp(ik_xx) I_{\nu;p_y}(x)
\end{equation}
With the substitution $X=\sqrt{q_eB}x$ and $C=p_{4y}/\sqrt{q_eB}$, the
above Fourier transform reduces to
\begin{eqnarray}
&&\int_\infty^\infty \frac{dX}{\sqrt{q_eB}}\exp (ik_xx)\left (
\frac{q_eB}{\pi} \right )^{1/4} \frac{1}{\sqrt{\nu!} 2^{\nu/2}} 
\nonumber \\ && \exp \left [ - \frac{1}{2} (X-C)^2\right ] H_\nu (X-C) 
\nonumber \\ &=& \frac{1}{(q_eB)^{1/4} \sqrt{\nu !} 2^{(\nu-1)/2}} i^\nu
H_\nu(k_x)  \nonumber \\ && \exp \left ( \frac{iCk_x}{\sqrt{q_eB}}
-\frac{k_x^2}{2q_eB} \right ) 
\end{eqnarray}
where $k_x=p_{1x}-p_{2x}-p_{3x}$. Then we have after some algebraic
manipulation
\begin{equation}
u_e^{\uparrow}=\sqrt{\frac{\varepsilon_4+m_e}{2\varepsilon_4}}
\left( \begin{array}{c} C_1H_\nu(k_x)\\ 0\\
C_3H_\nu(k_x)\\ C_4H_{\nu-1}(k_x)  
\end{array} \right)
\end{equation}
where
\begin{equation}
C_1=\frac{1}{(q_eB)^{1/4}\sqrt{\nu!}2^{(\nu-1)/2}}i^\nu \exp \left (
\frac{iCk_x}{\sqrt{q_eB}}-\frac{k_x^2}{2q_eB} \right ),
\end{equation}
\begin{equation}
C_3=\frac{p_{4z}}{\varepsilon_4+m_e}C_1
\end{equation}
and
\begin{eqnarray}
C_4 &=&-\frac{\sqrt{q_eB}2\nu}{(\varepsilon_4+m_e)
(q_eB)^{1/4}\sqrt{\nu!}2^{(\nu-1)/2}}i^\nu \nonumber \\ && \exp \left (
\frac{iCk_x}{\sqrt{q_eB}}-\frac{k_x^2}{2q_eB} \right )
\end{eqnarray}
Similarly the down spin state is given by
\begin{equation}
u_e^{\downarrow}=\sqrt{\frac{\varepsilon_4+m_e}{2\varepsilon_4}}
\left( \begin{array}{c} 0 \\C_2^\prime H_{\nu-1}(k_x)\\
C_3^\prime H_\nu(k_x)\\ C_4^\prime H_{\nu-1}(k_x)  
\end{array} \right)
\end{equation}
where
\begin{equation}
C_2^\prime =\frac{1}{(q_eB)^{1/4}\sqrt{(\nu-1)!}2^{(\nu-2)/2}}i^{\nu-1}
\exp \left (
\frac{iCk_x}{\sqrt{q_eB}}-\frac{k_x^2}{2q_eB} \right ),
\end{equation}
\begin{eqnarray}
C_3^\prime &=&-\frac{\sqrt{q_eB}}{(\varepsilon_4+m_e)
(q_eB)^{1/4}\sqrt{(\nu-1)!}2^{(\nu-2)/2}}i^{\nu-1} \nonumber \\ && \exp \left (
\frac{iCk_x}{\sqrt{q_eB}}-\frac{k_x^2}{2q_eB} \right )
\end{eqnarray}
and
\begin{equation}
C_4^\prime =\frac{p_{4z}}{\varepsilon_4+m_e}C_2^\prime
\end{equation}
Now by some rearrangement, integration over $p_{1x}$ can very easily be
performed and is given by $\sqrt{\pi q_e B}$. Whereas, the
integrations over $p_{1y}$, $p_{1z}$ and $d^3p_2$ can  be
evaluated trivially with the help of delta functions. Then finally,  we have 
after
substituting $(\mu_u-\epsilon_3)/T=x_u$ and $(\mu_e-\epsilon_4)/T=x_e$,
the rate for the process (1)
\begin{eqnarray}
\Gamma_1&=& \frac{3G^2(q_eB)}{2\pi^6} T^4 \cos^2 \theta_c \mu_u \mu_e
p_{Fu}\sum_{\nu_e=0}^{[\nu_e^{\rm{max}}]} \left ( \frac{1}{p_{Fe}}
\right ) \nonumber \\
&&\int_{-\infty}^\infty \left ( x_u+ x_e -\frac{\mu_u+\mu_e-\mu_d)}{T}
\right )^2 f(x_u)f(x_e) dx_u dx_e 
\end{eqnarray}
Similarly, the rate for the process (2) is given by 
\begin{eqnarray}
\Gamma_2&=& \frac{3G^2(q_eB)}{2\pi^6} T^4 \cos^2 \theta_c \mu_u \mu_e
p_{Fu}\sum_{\nu_e=0}^{[\nu_e^{\rm{max}}]} \left ( \frac{1}{p_{Fe}}
\right ) \nonumber \\
&&\int_{-\infty}^\infty \left ( x_u+ x_e +\frac{\mu_u+\mu_e-\mu_d)}{T}
\right )^2 f(x_u)f(x_e) dx_u dx_e 
\end{eqnarray}
In the above expressions, $p_{Fe}=(\mu_e^2-m_e^2-2\nu_eq_eB)^{1/2}$ is
the electron Fermi momentum.
Then as mentioned before, the rates of the processes (3) and (4) are
obtained from $\Gamma_1$ and $\Gamma_2$ respectively.
Whereas, the rates for both the direct
and reverse process as shown by reaction (5) are given by the zero field
values \cite{R13}.

Now knowing these rates we have solved the kinetic equations for
magnetic field strength $B=10^{14}$Gauss. The time evolution of the
fractional abundances for various components is shown in fig.(1).
\begin{figure} 
\psfig{figure=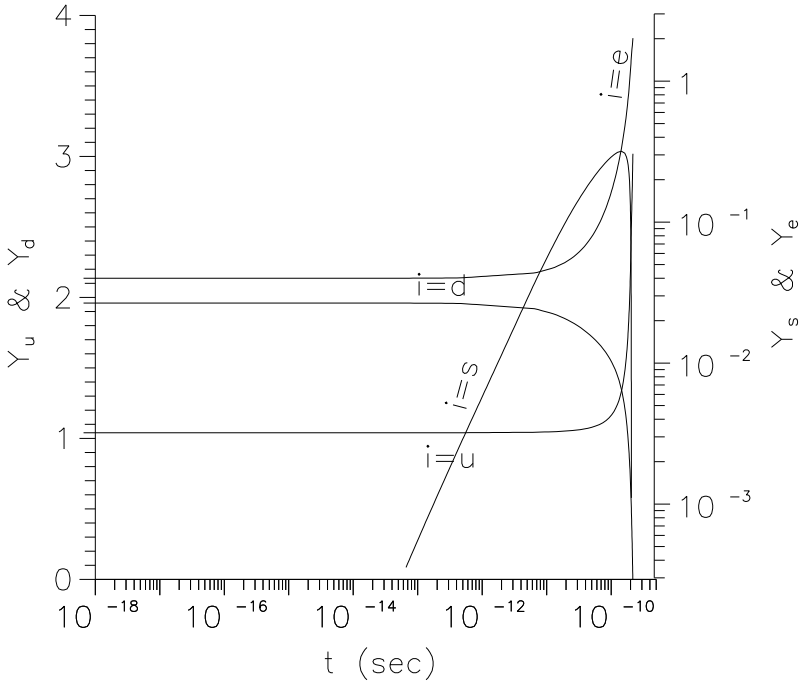,height=0.5\linewidth}
\caption[]{ Fractional abundances for various species  when only electrons 
are affected by quantizing 
magnetic field ($B=10^{14}$G). }
\end{figure}

This figure shows that in $\beta$-equilibrium condition there are mainly
$u$-quarks and electrons in the quark matter system. Then from 
dimensionality of the extended phase
space one can easily visualize  that the system is energetically much
more unstable than neutron matter of identical physical condition.  We
have also checked the result from explicit free energy calculation for
various core densities. 

Hence we conclude that stable quark
matter phase can not exist at the core of a newborn neutron star if the
magnetic field strength exceeds the critical value $\sim 4.4\times
10^{13}$Gauss. In fact,  we can now
very strongly demand that  at the core of a young neutron star  even with
moderately strong magnetic field, quark matter can not exist. Hence we
expect that is also possible to
extrapolate this conclusion to the quark matter system when all the 
charged components are affected, but the field strength is not high enough to
fill only the zeroth Landau levels. Because of mathematical difficulty,
of course, we are unable to show it explicitly. Finally,
we do believe, that if some system is energetically unstable, the
nature will not allow its creation at the very beginning. Therefore, the
possibility of quark-hadron phase transition at the core of a strongly
magnetized young neutron star is an open question.

\noindent Acknowledgment: The author is thankful to the Department of 
Science and Technology, Govt. of India, for partial financial support to 
this work, Sanction number:SP/S2/K3/97(PRU).  
\end{document}